\documentclass[prl,twocolumn,showpacs,floatfix]{revtex4}

\usepackage{amsfonts}
\usepackage{graphicx}

\begin{document}

\author{K. Shiokawa$^{1,2}$}
\author{D.A. Lidar$^1$}
\affiliation{$^1$Chemical Physics Theory Group, Chemistry Department,
University of Toronto, 80 St. George St., Toronto, Ontario M5S 3H6, Canada \\
$^{2}$Department of Physics, University of
Maryland, College Park, MD 20742, USA}
\title{Dynamical Decoupling Using Slow Pulses: Efficient Suppression of $1/f$
Noise}

\title{Dynamical Decoupling Using Slow Pulses: Efficient Suppression of $1/f$
Noise}

\begin{abstract}
The application of dynamical decoupling pulses to a single qubit
interacting with a linear harmonic oscillator bath with $1/f$
spectral density is studied, and compared to the Ohmic
case. Decoupling pulses that are slower than the fastest bath time-scale
are shown to drastically reduce the
decoherence rate in the $1/f$ case. Contrary to conclusions drawn from
previous studies, this shows that dynamical decoupling pulses do not
always have to be ultra-fast. Our results explain a recent experiment
in which dephasing due to $1/f$ charge noise affecting a charge qubit
in a small superconducting electrode was successfully suppressed using
spin-echo-type gate-voltage pulses.
\end{abstract}

\pacs{03.67.Hk,03.65.-w,03.67.-a,05.30.-d}

\maketitle
The most serious problem in the physical implementation
of quantum information processing is that of maintaining quantum
coherence. Decoherence due to interaction with the environment can
spoil the advantage of quantum algorithms \cite{Unruh:95}. One of
the proposed remedies is the method of ``dynamical decoupling'',
or ``bang-bang'' (BB) pulses, in which strong and sufficiently
fast pulses are applied to the system. In this manner one can
either eliminate or symmetrize the system-bath Hamiltonian so that
system and bath are effectively decoupled
\cite{Viola:98,Uchiyama:02,Agarwal:01,Search:00,Zanardi:98bViola:00aViola:01aWuLidar:01bWuByrdLidar:02ByrdLidar:01aViola:02,Duan:98e,Viola:99,Vitali:99Vitali:01,Facchi:03,LidarWu:02}.
The BB method was proposed in \cite{Viola:98}, where a
quantitative analysis was first performed for pure dephasing in
the linear spin-boson model: $H_{SB}=g \sigma _{z}\otimes B$,
where $\sigma _{z}$ is the Pauli-$z$ matrix and $B$ is a Hermitian
boson operator. The analysis was recently extended to the
non-linear spin-boson model, with similar conclusions about
performance \cite{Uchiyama:02}. Decoupling also has been applied
to the suppression of spontaneous emission \cite{Agarwal:01} and
magnetic state decoherence induced by collisions in a vapor
\cite{Search:00}. Since the decoupling pulses are {\em strong} one
ignores the evolution under $H_{SB}$ while the pulses are on, and
since the pulses are {\em fast} one ignores the evolution of the
bath under its free Hamiltonian $H_{B}$ during the pulse cycle.
The latter assumption is usually stated as:
\begin{equation}
\Delta t\ll 1/\Lambda_{UV} ,
\label{eq:BBcond}
\end{equation}
where $\Delta t$ is the pulse interval length and $\Lambda _{UV}$
is the high-frequency cutoff of the bath spectral density $I(\omega )$
\cite{Viola:98} [see Eq.~(\ref{eq:Iw}) below]. It can be shown that the overall system-bath coupling strength
$g$ is then renormalized by a factor $\Delta t \Lambda
_{UV}$ after a cycle of decoupling pulses \cite{Viola:99}, and
that the bath-induced error rate is
reduced by a factor proportional to $(\Delta t \Lambda _{UV})^{2}$ \cite{Duan:98e}. A temperature $T>0$ sets an additional, thermal decoherence time
scale that must be beat in order for the decoupling method to work \cite{Viola:98,Vitali:99Vitali:01}.

The conclusion~(\ref{eq:BBcond}) is extremely stringent, as the
timescale $\Delta t$ may be too small to be practically
attainable. Moreover, as we show below, and has been argued before on
the basis of the inverse quantum Zeno effect \cite{Facchi:03}, decoherence may be enhanced, rather than
suppressed, if (\ref{eq:BBcond}) is not satisfied. Eq.~(\ref{eq:BBcond}) is based on studies in which the
bath was modeled as a system of harmonic oscillators, with a
spectral density of the form $I(\omega )\propto \omega ^{\nu}e^{-\omega /\Lambda _{UV}}$,
with $\nu>0$ \cite{Viola:98}, or using a flat spectral density
with a finite cutoff $\Lambda _{UV}$
\cite{Vitali:99Vitali:01}, or without reference to a specific
spectral density but emphasizing features of its high-frequency
components \cite{Duan:98e,Uchiyama:02}. However, a ubiquitous
class of baths does not fall into this category, and we show here
that then the condition~(\ref{eq:BBcond}) is {\em overly
restrictive}. This is the case for so-called $1/f$ noise, or more
generally $1/f^{\alpha }$ ($ \alpha >0$). In these cases the bath
spectral density decays as a power law, bounded between
infrared (IR, lower) and ultraviolet (UV, upper) cutoffs $\Lambda_{IR}$ and $\Lambda_{UV }$, respectively. In quantum computer
implementations this is often attributable to (but certainly not
limited to) charge fluctuations in electrodes providing control
voltages. The need for such electrodes is widespread in quantum
computer proposals, e.g., trapped ions (where observed $1/f$ noise
was reported in \cite{Turchette:00}), quantum dots
\cite{Burkard:99Levy:01a}, doped silicon \cite{Kane:98Vrijen:00},
electrons on helium \cite{Platzman:99}, and superconducting qubits
\cite{Nakamura:02}. In the latter case, in a recent experiment
involving a charge qubit in a small superconducting electrode
(Cooper-pair box), a spin-echo-type version of BB was
successfully used to suppress low-frequency energy-level
fluctuations (causing dephasing) due to $1/f$ charge noise
\cite{Nakamura:02}. Here we explain the origin of such a result
and discuss its general applicability.

On the time scale $t > 1/\Lambda _{UV}$, the details of the
system-bath interaction and internal bath dynamics become
important. These details are captured by the bath spectral density $I(\omega)$.
Since for $1/f$ noise most of this density is concentrated in the low, rather than the high-end of the
frequency range, it turns out that in this case BB with slow pulses ($\Delta t >
1/\Lambda _{UV}$) depends
more sensitively on the lower than on the upper cutoff. In
particular, we show that the suppression of dephasing is more
effective when the noise originates in a bath with $1/f$ spectrum
than in the Ohmic case [$\nu =1$ in Eq.~(\ref{eq:Iw})], owing to the abundance of IR modes
in the former. In the following we present the
results of our analysis contrasting BB for $1/f$ and Ohmic baths.

{\it Decoupling for spin-boson model}.--- We consider the linear
spin-boson model including periodic decoupling pulses. We first
briefly review and somewhat simplify the results derived in
\cite{Viola:98}. We use $ k_{B}=\hbar =1$ units.
The Hamiltonian is
\begin{eqnarray*}
H&=&H_{S}+H_{B}+H_{SB}+H_{P} \\
 &=&{\frac{\epsilon }{2}}\sigma
_{z}\,+\sum_{k}\,\omega _{k}b_{k}^{\dagger }b_{k}+\sum_{k}\,\sigma
_{z}(g_{k}^{\ast }b_{k}+g_{k}b_{k}^{\dagger })+H_{P}\;,
\end{eqnarray*}
where the first (second) term governs the free system (bath) evolution; the
third term is the (linear) system-bath interaction in which $b_{k}$ is the $k$th-mode boson annihilation operator and $g_{k}$ is a coupling constant; and the
last term is the fully controllable Hamiltonian generating the decoupling
pulses:
\[
H_{P}(t)=\sum_{n=1}^{N}\,V_{n}(t)\,
e^{i\epsilon t \sigma _{z}/2} \sigma _{x} e^{-i\epsilon t \sigma _{z}/2}\;,
\]
where the pulse amplitude $V_{n}(t)=V$ for $t_{n}\leq t\leq t_{n}+\tau$ and $0$ otherwise, lasting for a duration $\tau \ll \Delta t,$ with $
t_{n}=n\Delta t$ being the time at which the $n$th pulse is
applied. The properties of the bath are captured by its spectral density
\begin{equation}
I(\omega )=\sum_{k}\delta (\omega -\omega _{k})|g_{k}|^{2}.
\label{eq:Iw}
\end{equation}

The reduced system density matrix is obtained from the total density matrix
by tracing over the bath degrees of freedom
\[
\rho _{S}(t)={\rm Tr}_{B}\left[ \rho (t)\right] ={\rm Tr}_{B}\left[ U(t)\rho
  _{S}(0)\otimes \rho _{B}(0)U^{\dagger }(t)\right] \;,
\]
where we have assumed a factorized initial condition between the
system and thermal bath, and $U(t)$ is the time evolution
generated by $H$: $U(t)= {\cal T} \exp \left[
-i\int_{0}^{t}ds\,H(s)\right]$ (${\cal T}$ denotes time ordering).
We are interested in how decoupling improves the system coherence,
defined as $\rho _{01}(t)=\langle 0|\,\rho _{S}(t)\,|1\rangle$.
In the interaction picture with respect to $H_{S}$ and $H_{B}$ the
result in the absence of decoupling pulses (free evolution) is:
$\rho _{01}^{I}(t)=e^{-\Gamma _{0}(t)}\rho _{01}^{I}(0)$, where
\begin{equation}
  \Gamma _{0}(t)= \int_{\Lambda_{IR}}^{\Lambda_{UV }} d\omega \,\coth \left(
\frac{\beta \omega }{2}\right) \, {\frac{1-\cos \omega
t_{2N}}{\omega ^{2}}} I(\omega )\nonumber
\label{eq:gamma0}
\end{equation}
$\beta =1/(k_{B}T)$.
In the Schr\"{o}dinger picture there are oscillations at the natural
frequency $\epsilon $, i.e., $\rho _{01}(t)=e^{-i\epsilon t}\rho _{01}^{I}(t)$.

Similarly in the presence of the decoupling pulses, at
$t_{2N}=2N\Delta t$, $\rho _{01}^{I}(t_{2N})=e^{-i\epsilon t_{2N}}e^{-\Gamma
  _{P}(N,\Delta t)}\,\rho _{01}^{I}(0)$,
where we can show from Eqs.~(46),(47) of \cite{Viola:98} that
\begin{eqnarray}
  \Gamma _{P}(N,\Delta t)&=& \label{eq:D1overf} \\
  4\int_{\Lambda_{IR}}^{\Lambda_{UV }} d\omega \,&\coth& \left(
\frac{\beta \omega }{2}\right) \, {\frac{1-\cos \omega
t_{2N}}{\omega ^{2}}}\,I(\omega )\,\tan ^{2}\left( \frac{\omega \Delta
t}{2}\right) \nonumber
\end{eqnarray}
The $\tan ^{2}\left( \frac{\omega \Delta t}{2}\right) $ term (which
was not found in \cite{Viola:98}) is
the suppression factor arising from the decoupling procedure. In
effect, the bath spectral density in the presence of decoupling pulses
has been transformed from $I(\omega)$ to the singular spectral density
$I'(\omega)=I(\omega)\tan
^{2}\left( \frac{\omega \Delta t}{2}\right)$. Note, however, that the
singularity of $\tan^2$ at $\omega \Delta t = (2n+1) \pi$ for an integer $n$
is canceled by the vanishing of $1-\cos \omega t_{2N}$ at the same points,
so $\Gamma _{P}$ remains
finite. Nevertheless, and as already pointed out in \cite{Viola:98},
the value $\omega \Delta t = \pi$ is special: In the limit $N\gg 1$
the integrand of Eq.~(\ref{eq:D1overf}) is highly oscillatory for
$\omega \Delta t > \pi$, and grows to $16N^2$ at $\omega \Delta t =
\pi$. Thus, decoherence suppression is effective when
\begin{equation}
  \Lambda_{UV }\Delta t<\pi .
\label{eq:constraint}
\end{equation}
This is an upper bound on $\Delta t$ that is independent of the
specific form of $I(\omega)$. Note further that decoupling
\emph{enhances} decoherence from all modes with $(4n+1)\pi/2 <
\omega \Delta t < (4n+3)\pi/2$, since for these values
$\tan^2(\omega \Delta t/2)>1$. However, this effect may be
quenched if the weight of these modes is sufficiently low; this is
indeed what happens in the $1/f$ case.

{\it Results for $1/f$ and Ohmic spectral densities}.---
Let us now assume that the spectral density has the following form:
\begin{equation}
I(\omega )=\gamma \omega ^{\nu },\quad \nu =\pm 1,
\end{equation}
with UV cutoff $\Lambda_{UV }$ and IR cutoff
$\Lambda_{IR}$. Thus we are comparing $1/f$ noise (the case $\nu
=-1$) to an Ohmic bath (the case $\nu =1$, considered in
\cite{Viola:98}).

\begin{figure}
  \includegraphics[height=5cm,angle=0]{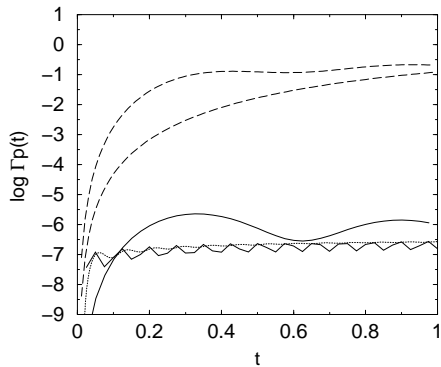}
\caption{Temporal behavior of the logarithm of the decoherence
factors at $T=0$. The initial coherence
$\protect\rho^{I}_{01}(0)=1$. Parameters are: $\gamma=0.05$,
$\Lambda_{UV}=10$ for Ohmic and $\gamma=0.25$,
$\Lambda_{UV}=80$ for $1/f$, $\Lambda_{IR}=1,\Delta t=0.025$
for both. Thick solid (dashed) line: $1/f$ case with (without)
decoupling pulses. Thin solid (dashed) line: Ohmic case with
(without) decoupling pulses. Eq.~(\ref{eq:gamma0}) was used for
the case without decoupling pulses, while Eq.~(\ref{eq:D1overf})
was used for the case with decoupling pulses at each $t=t_{2N}$.
The dotted line is our analytical result in Eq.~(\ref{eq:Gp0}).}
\label{fig1}
\end{figure}

To explain the effect of pulses qualitatively, we approximate
$\tan^2x$ by $x^2 (1- 2x/\pi)^{-1}$, which allows us to obtain an
explicit form for $\Gamma_P$ for $0 \le \Lambda_{UV }\Delta t <
\pi/2$. We further expand $\coth x \approx 1+2\exp(-2x)$ ($x>1$).
Then, the contribution to $\Gamma _{P}$ for $1/f$ noise at low
temperature is the sum of the zero temperature part
\begin{widetext}
\begin{eqnarray}
\Gamma _{P}^{(T=0)}(N,\Delta t) &=&
 \gamma \left( \Delta t\right)^{2} \left[
\log \left\{ {\frac{ \Lambda _{UV} }{ \Lambda _{0} }}\right\}
-\log \left\{
 \frac{ \pi - \Lambda _{UV} \Delta t }
{ \pi - \Lambda_{IR} \Delta t } \right\}
-{\rm Ci} \left( \Lambda _{UV
}t_{2N}\right) +{\rm Ci}\left( \Lambda _{0}t_{2N}\right) +
O(\Delta t) \right] \;, \label{eq:Gp0}
\end{eqnarray}
and the low temperature correction
\begin{eqnarray}
\Gamma_{P}^{(T > 0)}(N,\Delta t)  &=&
 {\frac{\gamma \left( \Delta t\right)
^{2}}{2}}
 \left[ \log \left( 1+ T^2 t_{2N}^2
 \right)
+  \frac{2 \Delta t T }{\pi }
 \left\{ 1 - \frac{1}{ 1+ T^2 t_{2N}^2 } \right\}
  + O(T^2)
 \right],
\label{eq:GpT}
\end{eqnarray}
\end{widetext}
where ${\rm Ci}$ (${\rm Si}$) is the cosine (sine) integral.
  In Eq.~(\ref{eq:GpT}), the limits
$\Lambda_{IR} \rightarrow 0$ and $\Lambda_{UV}\rightarrow
\infty$ are taken. All terms are finite in these limits.
 The first
and second terms in $\Gamma^{(T=0)}_{P}(N,\Delta t)$ (independent of $t_{2N}$) determine the
asymptotic value $\Gamma^{(T=0)}_{P}(\infty,\Delta t)$; the remainder is a
damped oscillatory part, given by the difference of two cosine
integrals, that vanishes at long times. The second logarithmic
term diverges as the pulse interval approaches the inverse UV
cutoff frequency time scale of the bath leading to
decoherence enhancement from the $\tan^{2}$ term in
Eq.~(\ref{eq:D1overf}).
These behaviors are reflected in
 the exact solutions displayed in
 Fig.~\ref{fig1}.
The leading order finite temperature correction $\Gamma^{(T>0)}_{P}(N,\Delta t)$
can be separated into two terms.
The first term characterizes the
asymptotic power law decay and the second term gives the initial
damping and the asymptotic relaxation to the $t_{2N}$-independent constant.

In Fig.~\ref{fig1} the logarithm of the
decoherence factors $\Gamma _{0}(t)$ (free evolution) and $\Gamma
_{P}(t)$ (pulsed evolution) for the $1/f$ and Ohmic cases are
shown. The smaller is $\Gamma $, the more coherent is the
evolution.  The apparent oscillations with a frequency given by $\Lambda
_{UV}$ are caused by the use of a sudden cutoff. Given the
parameters used in Fig.~\ref{fig1}, the standard timescale condition $\Delta t\ll 1/\Lambda
_{UV }$ is {\em not} satisfied in the $1/f$ case, while it is
($\Delta t \Lambda _{UV } = 0.25$) in the Ohmic case. The most
striking feature apparent in Fig.~\ref{fig1} is the highly
efficient suppression of decoherence in the case of $1/f$ noise,
in spite of the seemingly unfavorable pulse interval length. In
addition, it can be shown that
decoherence due to the $1/f$ bath is accelerated when the IR cutoff is
decreased, and is more sensitive to the IR
cutoff than the Ohmic case. This is a direct consequence of the
fact that most of the modes in $1/f$ spectrum are concentrated
around $\Lambda_{IR}$. For $1/f$ baths we therefore expect
slow and strong decoherence on a long time scale, that may be
efficiently suppressed by relatively {\em slow} and strong pulses.
A similar conclusion should be applicable to the more general
class of baths with $ 1/f^{\alpha }$ spectral density, since there
too most of the bath spectral density is concentrated in the low
frequency range.

\begin{figure}
  \includegraphics[height=5cm,angle=0]{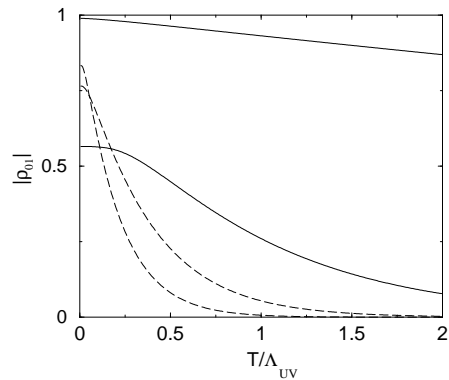}
  \caption{
The temperature dependence of coherence at $t=4$.
$\gamma=0.1$ for Ohmic case and $\gamma=0.5$ for $1/f$ case,
$\Lambda_{UV}=20$,
$\Lambda_{IR}=1$, $\Delta t=0.125$. Legend as in Fig.~\ref{fig1}.
}
  \label{fig3}
\end{figure}

\begin{figure}
  \includegraphics[height=5cm,angle=0]{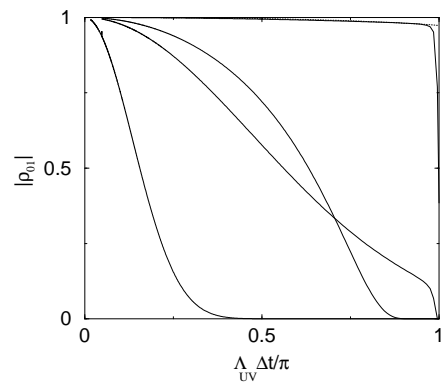}
  \caption{Coherence as a function of a pulse interval
  at finite temperature is plotted at $t=2$.
 Parameters are: $\gamma=0.5$,
$\Lambda_{UV}=100$, $\Lambda_{IR}=0.01$. Thick (thin) curves are
 $1/f$ (Ohmic) case.
$T=10$ for upper lines and $T=1000$ for lower lines.
The dotted line is from in Eqs.~(\ref{eq:Gp0})
and (\ref{eq:GpT}).
 }
  \label{fig4}
\end{figure}

For our pure dephasing case at finite
temperature, there is the thermal time scale $t_{\beta}\equiv
T^{-1}$ at which thermal fluctuations start affecting the system's coherence. In particular, for $T \gg
\Lambda_{UV}$, decoherence is governed by the thermal
fluctuations.
In Fig.~\ref{fig3}, a finite temperature result is shown.
The decoupling pulses enhance the decoherence for the Ohmic bath even at
low temperatures, since for the parameters chosen the condition (1) is
not satisfied. On the other hand, decoherence suppression in the $1/f$ case is
highly effective. At high temperature, it has been argued on the basis
of the
Ohmic case,
that decoupling pulses faster than the thermal frequency $T$ are
required to suppress decoherence \cite{Vitali:99Vitali:01}. Once
again, the nature of the bath can qualitatively modify this conclusion.
Thus decoupling by relatively slow pulses that
obey the condition $\Lambda_{UV} \Delta t \sim 1$, can still be
effective for decoherence suppression at elevated
temperatures. However, as the temperature increases, the effective
spectrum shifts toward low frequencies, and at the same time,
the influence of the environment increases.
Overall, BB becomes ineffective irrespective of the type of bath.
This explains the
breakdown of decoherence suppression at $T=1000$ in Fig.~\ref{fig4}.
Note from the figure that the suppression of decoherence for the $1/f$ bath is more effective than
for the Ohmic bath throughout the whole temperature regime.

For too slow pulses, BB accelerates the decoherence
\cite{Facchi:03}. For the Ohmic bath, as the interval approaches
the threshold value~(\ref{eq:BBcond}) from below, there is a
crossover from decoherence suppression to decoherence enhancement,
as shown in Fig.~\ref{fig4}. For the $1/f$ bath, suppression is
still effective for longer pulse intervals as long as $\Delta t
\Lambda_{UV} < \pi$ is satisfied.

It is of interest to compare our results with the gate voltage
pulse experiment performed in \cite{Nakamura:02} in a Cooper-pair
box. The corresponding parameter values in Eq.~(\ref{eq:D1overf})
are: $\gamma = 2 E_C^2 \alpha^2 / e^2 \hbar^2$, with the Josephson
charging energy $E_C=122$ [$\mu$eV] and the constant $\alpha=(1.3
\times 10^{-3} e)^2$ determined by the noise measurement. To
achieve 90$\%$ decoherence suppression with $\Lambda_{IR}=100$ [Hz]
and $\Lambda_{UV}=10$ [GHz] at $k_B T=5$ [$\mu$eV], the pulse
interval $\Delta t \sim 0.25$ [ns] is required from our analysis
based on Eq. (\ref{eq:D1overf}) with $N=1$. Although the pulse
sequence of ~\cite{Nakamura:02} differs from ours (theirs is the
$\pi/2-\pi-\pi/2$ spin-echo sequence), they play essentially the
same role. Our $\Delta t$ value roughly agrees with their value,
$\Delta t \sim 0.5$ [ns], deduced from Fig.~2 in
\cite{Nakamura:02}. This agreement nicely illustrates the
experimental feasibility of BB in the case of $1/f$ noise. The
effectiveness of spin-echo type pulses in relation to
superconducting qubits was also recently discussed in
\cite{Martinis:03}.
The spin-boson model is appropriate
for the study of $1/f$ noise due to a large number of weakly coupled
background charges\cite{PFFF:03}.

{\it Conclusions}.--- We have shown that the speed requirement of
the decoupling method should be stated relative to the {\em type}
of bath spectral density, and not just in terms of its upper
cutoff (baths with bimodal distributions provide another example of
this \cite{Uchiyama:02}). Most significantly, our exact results have demonstrated
that BB can be expected to be highly effective
in suppressing decoherence due to the ubiquitous $1/f$ noise,
without having to satisfy the stringent time constraints that may
render the method overly difficult to implement in other
instances. We expect this to have significant implications, e.g.,
for suppression of noise due to charge fluctuations in electrodes
providing control voltages in quantum computation. Such a result
has already been obtained experimentally in a Cooper-pair box
experiment \cite{Nakamura:02}, and is predicted to apply to
trapped-ion quantum computation as well \cite{LidarWu:02}.


{\it Acknowledgments}.---
  The present study was sponsored by NSERC and the DARPA-QuIST program (managed
  by AFOSR under agreement No. F49620-01-1-0468) (to D.A.L.).

\end{document}